\documentclass[runningheads]{llncs}
\usepackage[T1]{fontenc}
\usepackage{graphicx,verbatim,animate}
\usepackage{hyperref}
\usepackage{color}

\usepackage{amsmath}
\usepackage{booktabs}
\usepackage{makecell}
\begin{document}
\title{Systole-Conditioned Generative Cardiac Motion}
%

\author{Shahar Zuler \and
Gal Lifshitz \and
Hadar Averbuch-Elor \and
Dan Raviv} 
\authorrunning{S. Zuler et al.} 

\institute{Tel Aviv University, Tel Aviv, Israel} 
    
\maketitle              
\begin{abstract}

\begin{figure}
    \centering
    \animategraphics[autoplay, loop, width=0.8\linewidth]{10}
    {figures/generated_transition_teaser/}{0}{9}
       \caption{\textbf{From Input End-Systole (ES) to Fully Annotated Synthetic Sample:} Given an input ES frame, our framework generates a corresponding end-diastole (ED) frame, and a corresponding dense 3D ground-truth (GT) flow field. Our model creates fully annotated synthetic samples for cardiac CT, addressing the lack of annotated data in this domain. \textbf{Note: To view this interactive animation, which includes a gradual transition between the input ES and the generated ED, please open it with Adobe Acrobat.}}
    \label{fig:teaser}
\end{figure}

Accurate motion estimation in cardiac computed tomography (CT) imaging is critical for assessing cardiac function and surgical planning. Data-driven methods have become the standard approach for dense motion estimation, but they rely on vast amounts of labeled data with dense ground-truth (GT) motion annotations, which are often unfeasible to obtain.
To address this limitation, we present a novel approach that synthesizes realistically looking pairs of cardiac CT frames enriched with dense 3D flow field annotations.

Our method leverages a conditional Variational Autoencoder (CVAE), which incorporates a novel multi-scale feature conditioning mechanism and is trained to generate 3D flow fields conditioned on a single CT frame. By applying the generated flow field to warp the given frame, we create pairs of frames that simulate realistic myocardium deformations across the cardiac cycle. 
These pairs serve as fully annotated data samples, providing optical flow GT annotations.
Our data generation pipeline could enable the training and validation of more complex and accurate myocardium motion models, allowing for substantially reducing reliance on manual annotations.

Our code, along with animated generated samples and additional material, is available on our project page: \url{https://shaharzuler.github.io/GenerativeCardiacMotion_Page}\label{project_page}. 


\end{abstract}
\section{Introduction}
Cardiovascular diseases remain a leading cause of morbidity worldwide, highlighting the need for accurate cardiac imaging and motion analysis for effective diagnosis, treatment planning, patient management, and surgical interventions. CT scans offer non-invasive 3D visualization of the heart's structure and function across different phases of the cardiac cycle, providing critical insights into myocardial motion. 
Consequently, motion estimation between corresponding CT scans has become a pivotal task in biomedical imaging, motivating the development of data-driven models to capture the complexity of cardiac motion. 

While other imaging modalities, such as ultrasound (US) and magnetic resonance imaging (MRI), offer some automated motion estimation capabilities, they are hindered by limitations such as planar perception, operator dependence, lower image quality (US), and insufficient through-plane resolution (MRI).

Deep learning has significantly advanced cardiac motion estimation \cite{duchateau2020machine} through specialized architectures \cite{morales2021deepstrain,Zul_CardioSpectrum_MICCAI2024,meng2022mulvimotion}, the incorporation of priors and constraints \cite{hanania2023pcmc,inbook}, and various other approaches to model myocardial motion. 
However, the lack of dense GT optical flow annotations for cardiac CT remains a major challenge. Some research has focused on unsupervised learning to bypass this limitation, but even these models require annotated datasets for validation, making the absence of GT a key obstacle for both training and evaluation. 
Another approach involves hand-crafted synthetic datasets, which use prior knowledge to simulate \cite{segars20104d,AjaFernndez2012AMP} or deform \cite{zuler2024synthetic} voxel positions based on cardiac motion dynamics. However, such methods often fail to capture subtle myocardial deformations and imaging artifacts, leading to domain gaps that limit their effectiveness.

The remarkable success of generative models in producing high-quality synthetic data \cite{rombach2022high,achiam2023gpt,ramesh2022hierarchical} has led to their adoption in biomedical imaging for tasks such as cross-modality translation \cite{DAYARATHNA2024103046}, sequence or video generation \cite{yoon2023sadm,Li_Endora_MICCAI2024} and more, improving model performance in segmentation \cite{9761478}, image enhancement \cite{Cui_7T_MICCAI2024,Xio_Contrast_MICCAI2024,Zho_HeartBeat_MICCAI2024,ARMANIOUS2020101684} and diagnostics \cite{jiang2023s2dgan,cahan2024xray2ctpagenerating3dctpa}.
However, their potential for generating synthetic data specifically tailored to myocardial motion estimation remains largely unexplored, though recent efforts \cite{unleashing} indicate a growing interest.

In this work, we present a novel framework that leverages a CVAE \cite{sohn2015learning} to generate synthetic dense 3D flow fields conditioned on single CT frames, resulting in realistic cardiac image pairs fully annotated with dense GT optical flow.

Conditional Generative Adversarial Networks (GANs) \cite{mirza2014conditional} and diffusion models \cite{rombach2022high} generate high-quality samples but require large annotated datasets and substantial computational resources. 
In contrast, CVAEs offer a more data-efficient alternative, producing smoother outputs that align with the smooth nature of flow fields, making them well-suited for synthesizing dense 3D motion data in data-scarce scenarios.

In contrast to standard approaches that rely on a single representation of the conditioning signal, we introduce a novel multi-scale feature conditioning mechanism that integrates features of varying resolutions learned from a feature pyramid network (FPN).
This multi-scale feature conditioning mechanism provides richer, more anatomically relevant context, ensuring realistic and physiologically plausible deformation fields.

We evaluate our method through reconstruction error measurements.
Our method generates plausible, anatomically consistent flow fields (see Figures \ref{fig:teaser} and \ref{fig:animated_gradual_transition}), smooth latent space representation, and diverse, physiologically realistic variations (see Fig. \ref{fig:latent_space_sampling}).

%
\section{Methods}
We utilize a VAE with a novel multi-scale conditioning mechanism to generate synthetic, realistic deformations conditioned on a single end-systole (ES) frame, which are then used to warp the corresponding conditioning frame.

\subsection{Conditional Variational Autoencoder}

\begin{figure}
    \centering
    \includegraphics[width=0.95\linewidth]{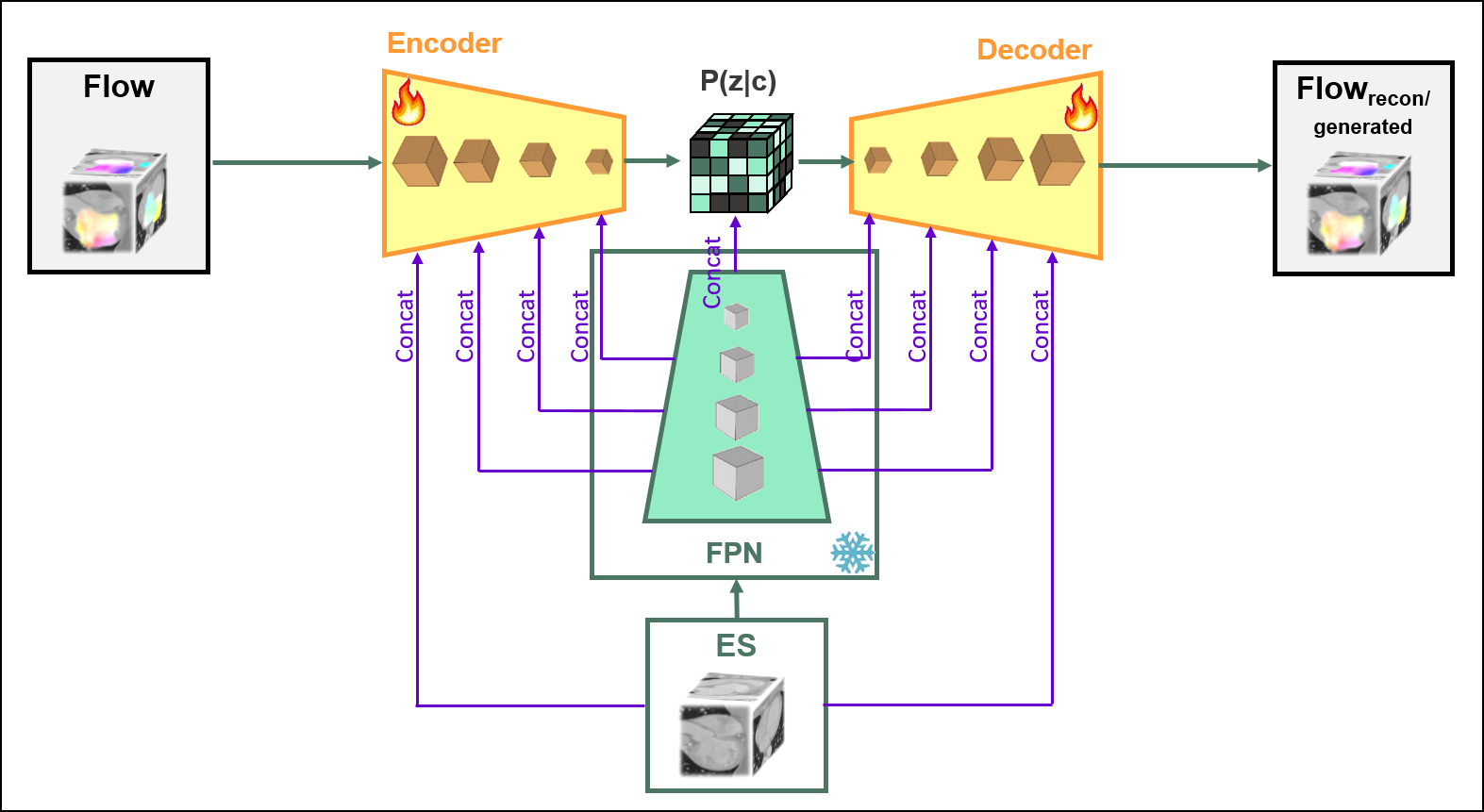}
    \caption{\textbf{Architecture of the CVAE Model Used for Flow Field Generation:} The encoder and decoder receive conditioning feature maps from a pretrained PWC-Net FPN. The decoder generates the flow field, which is used to warp the conditioning CT frame.}
    \label{fig:architecture}
\end{figure}

At the core of our method is a CVAE designed to generate 3D flow fields conditioned on single real CT frames. The architecture consists of an encoder, a decoder, and a multi-scale feature conditioning mechanism, as outlined in Fig. \ref{fig:architecture}.

\begin{itemize}
     
\item{\textbf{Encoder:}}  
The encoder structure is adapted from the FPN used in PWC-Net \cite{sun2018pwc}, extended to handle 3D flow fields. It takes concatenated inputs of the flow field \(X\) and conditioning frame \(C\), which are passed through \(L\) encoder levels. Each level consists of 3D convolutional layers.
This process compresses \(X\) into a latent representation \(Z\), conditioned on the input frame \(C\).

\item{\textbf{Decoder:}}  
The decoder mirrors the encoder structure, progressively upsampling the latent representation \(Z\) to the original resolution using transposed convolutions (deconvolutions). The final layer produces the reconstructed flow field.

\item{\textbf{Multi-Scale Feature Conditioning Mechanism:}}  
To condition the CVAE, we utilize an FPN derived from the optical flow network PWC-Net, adapted for 3D images. PWC-Net variants have been successfully employed in optical flow tasks in medical imaging \cite{lifshitz2021unsupervised,Zul_CardioSpectrum_MICCAI2024,tehrani2021mpwc}. The FPN is pre-trained as part of the PWC-Net framework for cardiac CT flow estimation, allowing it to extract multi-scale features from the conditioning frame \(C\) at each level, which are specifically tailored for cardiac flow estimation.

At each level of the encoder or decoder, the corresponding feature maps are concatenated with the level’s activations to retain hierarchically relevant conditioning information, which is then passed to the next layer.
\end{itemize}

Unlike conventional image generation tasks that condition solely on the image itself or a single embedding, our model leverages learned \textbf{multi-scale features} alongside the original image for conditioning. This hierarchical approach introduces a novel mechanism for conditioning on multi-scale optical flow features. The structural similarity between the encoder-decoder architecture of the CVAE and the FPN facilitates the concatenation of each condition level with it's corresponding layer in the encoder and the decoder, ensuring that both local and global information are incorporated.

The weights of the FPN are kept frozen during CVAE training, a common practice in conditional generative models, ensuring the conditioning features remain consistent throughout the learning process.

\subsection{Training Scheme}
\begin{figure}
    \centering
    \includegraphics[width=0.95\linewidth]{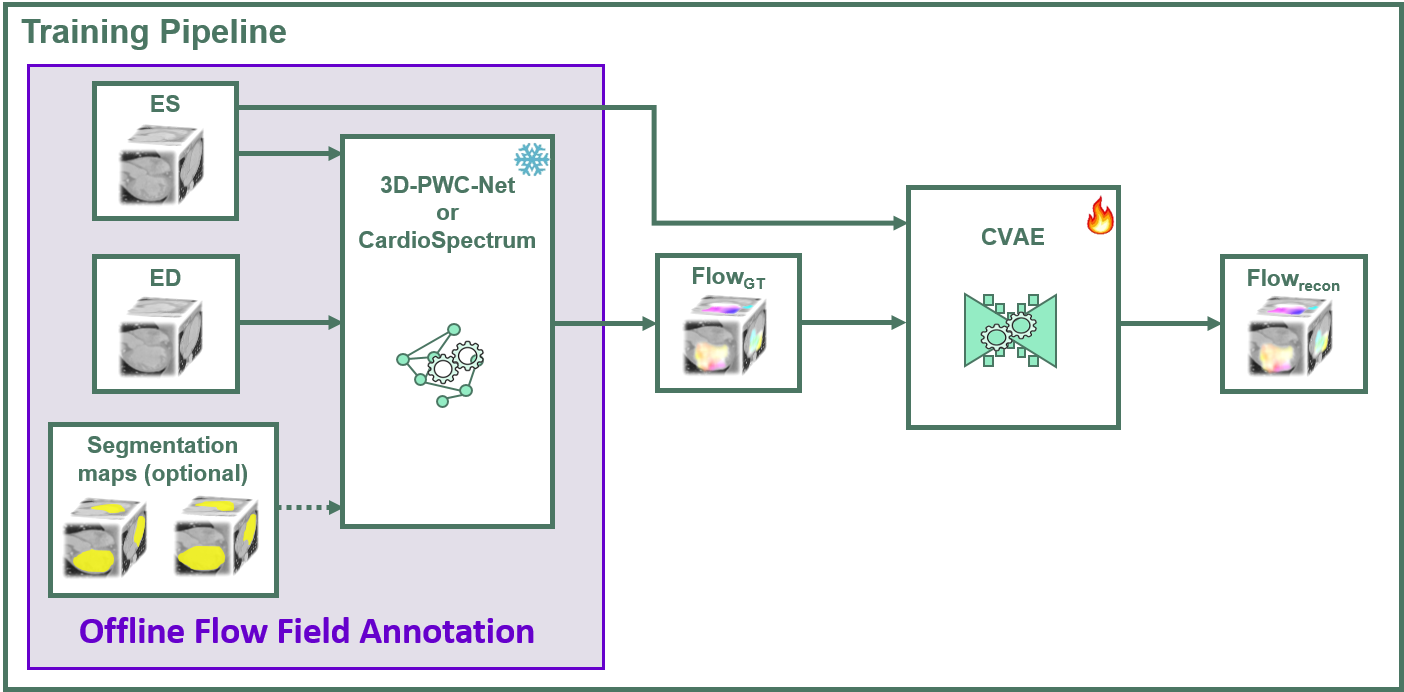}
        \caption{\textbf{Overview of the Training Pipeline:} ES and ED frames, along with optional segmentation maps, are used to compute GT flow fields via 3D-PWC-Net or CardioSpectrum. The CVAE then learns to reconstruct these flow fields, producing Flow\textsubscript{recon}.
        The offline process of generating \(Flow_{\text{GT}}\) annotations (shown in the purple box in the figure) occurs only once, prior to training.}
    \label{fig:training_pipeline}
\end{figure}

The CVAE is trained to reconstruct flow fields generated from pairs of ES and end-diastole (ED) frames, conditioned on the ES frame, as illustrated in Figure \ref{fig:training_pipeline}. GT flow fields are computed using established optical flow models, as described in \nameref{annotation_process}. 

The model minimizes a weighted combination of reconstruction loss (\(L_2\)) and Kullback-Leibler (KL) divergence \cite{kullback1951information}:

\begin{equation}
\mathcal{L}_{\text{total}} = \lambda_{\text{recon}} \cdot \mathcal{L}_{\text{recon}} + \lambda_{\text{KL}} \cdot \mathcal{L}_{\text{KL}}
\end{equation}

The weighting factors \(\lambda_{\text{recon}}\) and \(\lambda_{\text{KL}}\) are specified in Section \ref{implementation_details}.

\subsection{Sample Generation}  
\begin{figure}[h]
    \centering
    \includegraphics[width=0.95\linewidth]{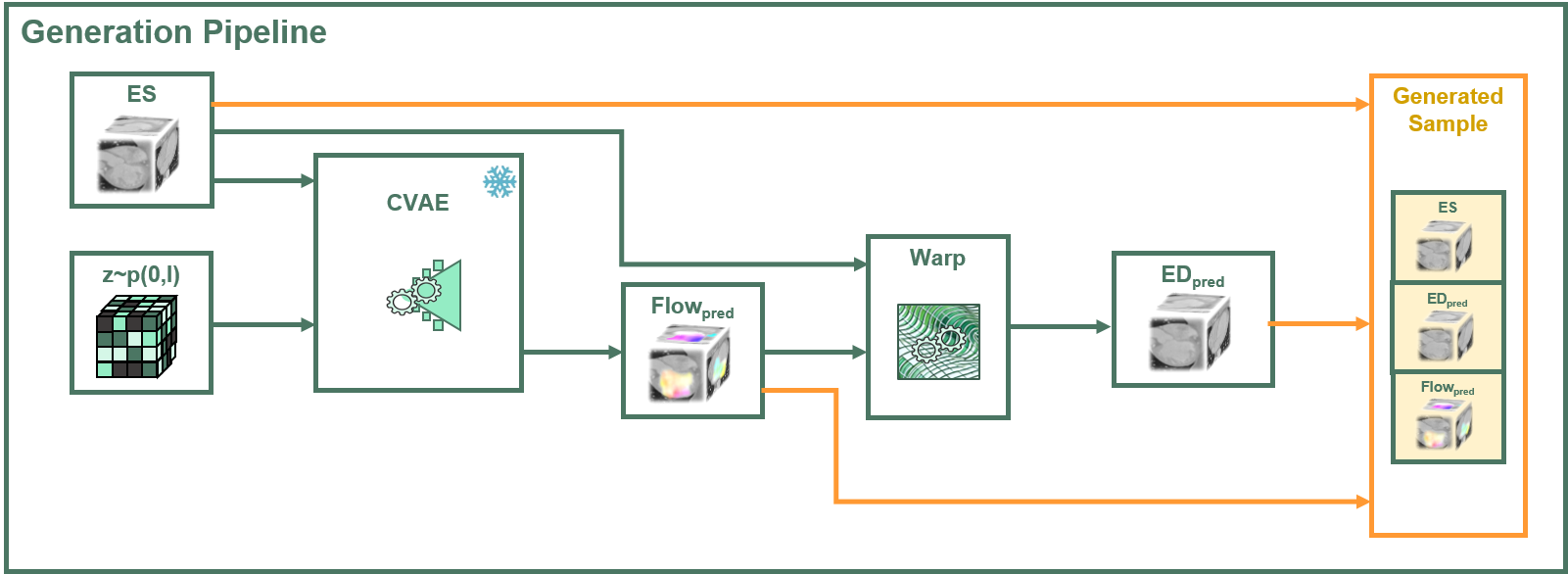}
    \caption{\textbf{Overview of the Generation Process:} The trained CVAE model generates a predicted flow field (Flow\textsubscript{pred}) from a conditioned ES frame and a random array \(Z\). This flow field is then used to warp the ES frame, producing an ED frame (ED\textsubscript{pred}), which, along with the flow, forms the final generated sample.}
    \label{fig:generation_pipeline}
\end{figure}

The generation pipeline produces synthetic annotated ES and ED pairs using the trained CVAE as illustrated in Figure \ref{fig:generation_pipeline}. 
Starting with an ES conditioning frame \(C\) and a latent vector \(Z\) sampled from a standard normal distribution, the decoder generates a flow field conditioned on \(C\) via concatenation.
This flow field warps the conditioning ES frame \(C\), producing a generated ED frame. 
Together, \(C\), the generated ED frame, and the generated flow field form a fully annotated synthetic sample.

By sampling different \(Z\) values, the CVAE produces diverse flow fields conditioned on the same \(C\), increasing variability in the synthetic dataset.

\begin{figure}
    \centering
    \animategraphics[autoplay, loop, width=0.8\linewidth]{10}
    {figures/generated_transition/generated42-}{0}{9}
       \caption{Visualization of a gradual transition between ES and the generated ED, shown across middle sections in three directions. \textbf{Note: To view this interactive animation, please open with Adobe Acrobat.}}
    \label{fig:animated_gradual_transition}
\end{figure}

\section{Experiments}
We conducted several experiments to evaluate the accuracy and realism of the generated flow fields.

The dataset consisted of 44 real cardiac 4DCT scans, including one from the MAGIX dataset sample via OsiriX \cite{osirix}, one from the 3D Slicer sample dataset \cite{slicer}, and 42 drawn from a private collection intended for research studies.

We used a leave-one-out cross-validation strategy, which enabled us to train the model on nearly the entire dataset while testing its generalizability on unseen samples.

\subsubsection{Flow Field Annotation Process} \label{annotation_process}
To generate GT flow fields for training the CVAE, we optimized known 3D optical flow models, and applied them to pairs of ES and ED frames extracted from the dataset.

For 40 scans, we employed a PWC-Net \cite{sun2018pwc}, a widely used optical flow estimation network, adapted for 3D images as outlined in \cite{lifshitz2021unsupervised}. For the remaining 4 scans, which included manually segmented left ventricle (LV) regions, we used CardioSpectrum \cite{Zul_CardioSpectrum_MICCAI2024} to estimate the flow fields.

\subsection{Implementation Details} \label{implementation_details}
We trained the CVAE for 2000 epochs with a batch size of 16 using the Adam optimizer and an a learning rate of $10^{-3}$. 
Our FPN consisted of \(L = 5\) levels.
The loss function combined reconstruction loss ($\lambda_{recon}=0.95$) and KL divergence loss ($\lambda_{KL}=0.05$).
Gradient clipping was used during training, with a maximum norm of 0.75.

To improve model robustness, data augmentations were applied, including random intensity noise (variance: 0.015), spatial shifts (maximum absolute shift: 10 pixels), and zoom transformations (range: [0.9, 1.1]).

\subsection{Reconstruction Evaluation}
During training, the Mean Square Error (MSE) between the reconstructed flow fields and the ground truth flow fields was computed. For 3D flow fields, this MSE can be interpreted as the mean Endpoint Error (\textbf{mEPE}), which measures the squared Euclidean distance between the predicted and true displacement vectors.

The average mEPE over the reconstructed flow fields in the leave-one-out cross-validation was \textbf{0.56} voxels, with a standard deviation of \textbf{0.58}, demonstrating the CVAE’s ability to generate flow fields closely aligned with ground truth annotations.

\begin{figure}
    \centering
    \includegraphics[width=0.95\linewidth]{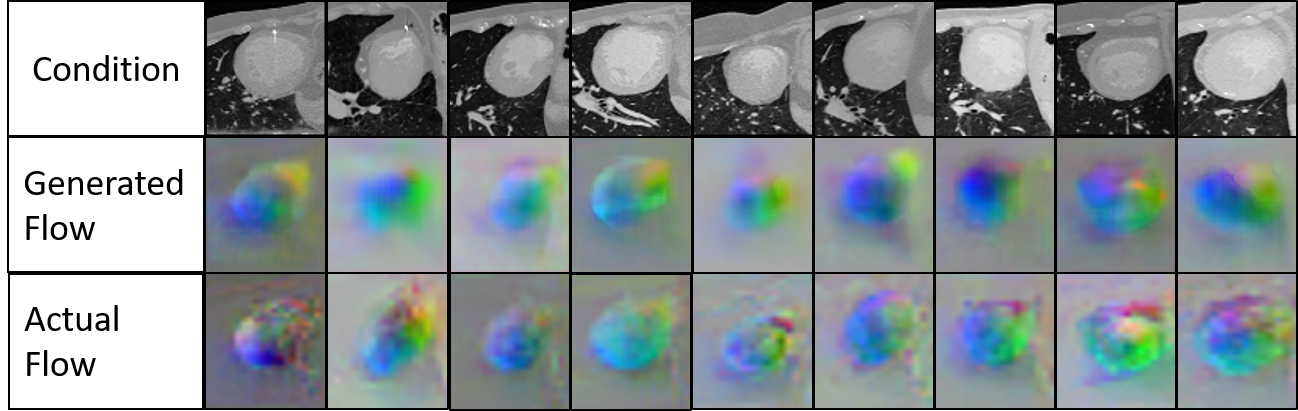}
       \caption{\textbf{Flow Field Comparison for Cardiac CT Conditions:} The top row ("Condition") shows 2D sections from validation CT scans (ES). The middle row ("Generated Flow") displays corresponding flow fields generated by the CVAE for each condition. The bottom row ("Actual Flow") presents the ground truth flow fields obtained from the original CT data. This comparison demonstrates the model's ability to generate plausible deformation patterns from real cardiac CT conditions.}
    \label{fig:generated_result}
\end{figure}

A qualitative illustration of these results is presented in Figure \ref{fig:generated_result}, which compares generated flow fields with their conditioning frames and GT flow fields.  The CVAE accurately captures anatomical structures and motion patterns, highlighting its capability to produce realistic and anatomically coherent deformations.

\begin{figure}
    \centering
    \includegraphics[width=0.99\linewidth]{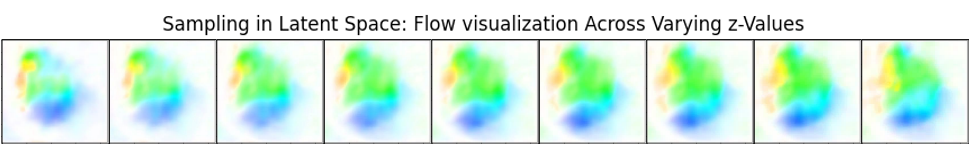}
    \caption{\textbf{Uniform Sampling Along an Arbitrary Line in Latent Space}: Flow fields generated by uniformly sampling latent $Z$ values along an arbitrary line in the latent space. The displayed sections are $\hat{x}$-direction slices, illustrating how the CVAE captures deformation variations for the same condition. 
    See the ancillary files \texttt{supp1.avi} and \texttt{supp2.avi} (available for download on the arXiv page) for dynamic video representations. 
    \texttt{supp1.avi} provides a sequential video of the frames shown here, while \texttt{supp2.avi} demonstrates transitions along a latent space plane using grid interpolation.}
    \label{fig:latent_space_sampling}
\end{figure}

\subsection{Conditional Latent Space Sampling}
To assess the CVAE’s ability to generate diverse flow fields, we visualized flow fields by varying the conditional latent variable $Z$ for a fixed conditioning frame. Figure \ref{fig:latent_space_sampling} illustrates uniform sampling along an arbitrary line in latent space, demonstrating a smooth and continuous latent space while also exhibiting notable variance, highlighting the model’s ability to generate diverse flow patterns.

\section{Conclusion}
In this work, we presented a novel, data-efficient framework based on CVAE for generating synthetic cardiac CT data. By leveraging multi-scale feature conditioning, our model generates anatomically realistic ES and ED frame pairs along with perfectly annotated 3D flow fields.

Quantitative evaluation demonstrated that the reconstructed flow fields achieved an average mEPE of \textbf{0.56} voxels, highlighting the accuracy of the reconstructed deformations. Qualitative results further validated the CVAE’s ability to capture complex myocardial motion while preserving anatomical coherence and generating diverse deformations through latent space sampling.

This approach addresses the critical challenge of limited annotated data in cardiac motion analysis by offering a practical and scalable alternative to manual annotation. The synthetic data generated by the CVAE can serve as a valuable resource for training and validating advanced models for myocardial motion estimation, paving the way for more robust and accurate cardiac CT analysis.

Future work will explore extending this framework to synthesize data for specific medical conditions and generating sequences that represent entire cardiac cycles, further enhancing its applicability and utility in medical imaging research.

%
%
%
\bibliographystyle{splncs04}
\bibliography{references}
%




\end{document}